# On the measurement by EDX
# of diffusion profiles of Ni/Cu assemblies

by

Olivier ARNOULD and François HILD[*]

LMT-Cachan

ENS de Cachan/CNRS/Université Paris 6

61, avenue du Président Wilson

F-94235 Cachan Cedex

France

[*]to whom correspondence should be addressed, Fax: +33 (0)1 47 40 22 40

E-mail: hild@lmt.ens-cachan.fr

**On the measurement by EDX of diffusion profiles of Ni/Cu assemblies**

Page Banner: EDX signal deconvolution

Authors:  Olivier Arnould and François Hild

LMT-Cachan

ENS de Cachan/CNRS/Université Paris 6

61, avenue du Président Wilson

F-94235 Cachan Cedex

France

Keywords:  EDX, pear-shaped emission volume, interdiffusion, concentration profiles, artefacts, deconvolution

Summary:

To characterise (inter)diffusion in materials, concentration profiles can be measured by EDX. It allows one to determine the chemical composition with a very good accuracy if measurement artefacts are accounted for. Standard phenomena (such as X-ray fluorescence) are usually corrected by commercial software. However, the effect of the pear-shaped volume of X-ray emission on the concentration profiles has to be considered. The paper describes the origin of this artefact, its consequences on measurements and will provide a practical solution (based on signal processing methods) to deconvolute the actual concentration profiles (or the diffusion coefficient) from the raw measurements.

Introduction:

**Scope of the study**

The evaluation of diffusion coefficients may require low scales of observation. Therefore practical studies are performed in an SEM. Concentration profiles are measured by an EDX technique that is based on the evaluation of the X-ray energies emitted by a sample impacted by the fast-moving primary electrons in the SEM. It allows one to identify and quantify the chemical compounds within a depth of about 1μm [1], which depends on the acceleration voltage of the primary electron beam and on the chemical composition. Phase identification and quantification depend on the accurate measurement of the value and intensity of the different peaks of the X-ray energy spectra. Quantitative determination is carried out by comparing the integrated intensities of selected peaks (*i.e.,* those who have the best number of counts, for a given time, like the $K_\alpha$ ones in our case) with standard X-ray data for reference materials [2]. These spectra can be measured at different locations of the sample surface (see the scan line in Fig. 1, when a Ni/Cu/Ni layered material of overall thickness $2(h_{Cu}+h_{Ni}) \approx 100$μm obtained by electrodeposition is analysed, where $2h_{Cu}$ denotes the thickness of the copper layer and $h_{Ni}$ that of each Nickel layer). Consequently, a concentration profile can be evaluated for the study of diffusion at a very low scale [3] thanks to a spatial resolution of 100nm and an energy resolution of about 130eV [4]. A mass concentration resolution less than 1% can be obtained in suitable conditions (*e.g.,* SEM chambers treated with Nitrogen gas to prevent it from outer pollution, clean EDX detector (Beryllium) window, optimal relative position of the latter with respect to the sample surface, optimal acceleration voltage for the studied materials, *e.g.,* three times the maximum peak energy of the spectrum, optimal tilt angle and working distance [4-5]).

**Measurement artefacts: a qualitative study**

Measurement artefacts must be considered if a very good accuracy is expected. Standard artefacts such as the spectrum background (Fig. 1) or the effect of too close peaks (*e.g.,* Ni(K$_\beta$) = 8.264keV and Cu(K$_\alpha$) = 8.047keV) are usually corrected by the software used to analyse the results of an EDX microprobe (*e.g.,* ZAF correction methods). However, in the case of diffusion, there is an artefact whose effects on the concentration profiles are very important and that is generally not treated by software. This artefact could be called *pear effect* and is caused by the shape of the volume of the X-ray emission (Fig. 2) whose dimensions can be of the same order of magnitude as the penetration length of diffusion (*i.e.,* the length over which interdiffusion has occurred). This effect can be qualitatively studied as shown in Fig. 2 when a half sphere of radius *r* models the emission volume.

As a first approximation, we assume that all points lying within the emission volume have the same efficiency of emission. If one wants to model the signal of Copper obtained when the emission volume traverses the Ni/Cu interface in the case of no diffusion, one has to calculate the part of the half sphere that lies in the Copper region since the measurement signal is proportional to its volume fraction (Fig. 3). The volume $V_{Cu}$ of the portion of sphere containing Copper evolves with the position of the beam axis *x* with respect to the concentration discontinuity located at $x = 0$

$$V_{Cu}(x) = \frac{\pi(x+r)^2}{3}\left(r - \frac{x}{2}\right) \quad \text{when } /x/ \leq r. \tag{1}$$

Figure 3 shows that the evolution of $V_{Cu}$ is very close to a normal cumulative distribution function, itself very similar to a diffusion profile in one-dimensional conditions [6]. Therefore, even though no diffusion occurred, the analysis of the raw measurements may conclude that some diffusion can be observed. Moreover, in the present case, the observed depths of diffusion are small [7] and this effect is all the more important since the true diffusion curve is then lost in this artefact. However, with the help of signal processing methods, it is possible to deconvolute the actual concentration profiles from the raw measurement when the real spatial X-ray emission distribution is identified. Furthermore, it must be noted that the rate of emission changes with the position in the pear-shaped emission volume, *i.e.,* the closer to the impact of the beam, the more emission of X-rays. This effect will be accounted for in the next section.

<u>Materials and Methods:</u>

**Theoretical deconvolution: a quantitative study**

Our work utilised a Hitachi S-510 SEM coupled with a PGT-EDX microprobe and IMIX software for imaging and X-ray signal processing. The raw measurements include the effect of the measurement spot (*i.e.,* the pear-shaped volume) on the actual diffusion curves. The measurement curve $M$ is then equal to the convolution between the real concentration profile $C$ by the emission function $E$

$$M(x) = \int\limits_{-\infty}^{+\infty} \int\limits_{-\infty}^{+\infty} \int\limits_{-\infty}^{0} C(\xi)\, E(x-\xi,\psi,\zeta)\, \mathrm{d}\xi\, \mathrm{d}\psi\, \mathrm{d}\zeta \qquad (2)$$

The spot in the two-dimensional case is assimilated to an axysimmetric normal probability density function $G_\sigma(\rho)$ centred at $\xi = x,\ \psi = 0\ (\rho = \sqrt{(x-\xi)^2 + \psi^2})$ and of standard deviation $\sigma$ since the farther the considered point at the material surface from the impact beam axis, the smaller the number of emitted X-rays for a given time. It is worth remembering that $\sigma$ mainly depends on the acceleration voltage of the primary electrons and on the atomic number of the impacted elements. Furthermore, the pear shape must be accounted for. It follows that a multiplicative function $P(\zeta)$ is used to describe the emission function $E$ (Fig. 4a). It is important to note here that the studied materials (*i.e.,* Nickel and Copper) have very close atomic numbers, therefore the variation of the dimension of the pear-shaped emission volume when crossing the interface is not significant. But for other material couples, one has to take that effect into account in the choice of the mathematical function representing the function $E$ (*e.g.*, with help of Monte-Carlo simulations [8, 9]). By noting that $G_\sigma(\rho) = G_\sigma(x-\xi)G_\sigma(\psi)$ and that $P$ is a unit function, *i.e.*, $\int_{-\infty}^{0} P(\zeta)\,\mathrm{d}\zeta = 1$, the measured value $M$ at the position $x$ is then defined by a *one-dimensional* convolution product

$$M(x) = \int_{-\infty}^{+\infty} C(\xi)\, G_\sigma(x-\xi)\, \mathrm{d}\xi = (C \otimes G_\sigma)(x) \tag{3}$$

where $\otimes$ denotes the convolution product. When no diffusion occurred, there is a concentration jump say at $\xi = 0$ (*i.e.*, $C(\xi) = 0$ if $\xi < 0$, $C(\xi) = 1$ if $\xi > 0$). It follows that a closed-form solution can be derived

$$M(x) = \int_{-\infty}^{x} G_\sigma(\chi)\, \mathrm{d}\chi = \frac{1}{2}\left[ 1 + \mathrm{erf}\left(\frac{x}{\sqrt{2}\,\sigma}\right) \right] \tag{4}$$

where erf denotes the error function. In the present case (see Fig. 1), two interfaces are present and located at $\xi = h_1$ and $\xi = h_2$ (*i.e.*, $C(\xi) = 0$ if $\xi < h_1$ or $\xi > h_2$, $C(\xi) = 1$ if $h_1 < \xi < h_2$). When no diffusion occurred, the previous results can be generalised and the raw measurement is expressed as

$$M(x) = \mathrm{erf}\left(\frac{x-h_1}{\sqrt{2}\,\sigma}\right) - \mathrm{erf}\left(\frac{x-h_2}{\sqrt{2}\,\sigma}\right) \tag{5}$$

Equation (4) is used to identify the standard deviation $\sigma$ on a reference material (here a bilayered material without diffusion, see Fig. 1, where the three unknowns are $\sigma$, $h_1$ and $h_2$). By using a least squares method, one gets $\sigma = 0.250\mu m$ (with an acceleration voltage of 15keV), $h_1 = 1.310\mu m$ and $h_2 = 4.005\mu m$ ($2h_{Cu} = h_2 - h_1 = 2.690\mu m$). It can be noted that an increment of $\pm 0.005\mu m$ from the optimal solution on any of the three unknowns leads to an increase of the order of 4% of the squared 2-norm of the residuals associated with the least squares method. Therefore, the last digit of the values of the unknowns takes only the value 0 or 5. Figure 4b shows a comparison between the experimental data and the least squares fit. A good agreement is obtained. It can be noted that the order of magnitude is consistent with the Monte-Carlo simulations dealing with Copper and shown in Fig. 2 [9].

**Identification of the diffusion coefficient**

We make the assumption that the diffusion coefficient $D$ is independent of the concentration. Consequently, the first Fick's law (*i.e.,* the flux of atoms is proportional to the concentration gradient) combined with the equation of continuity (*i.e.,* conservation law) yields the diffusion equation that can be solved, in any one dimensional case, by means of the Fourier transform. One can show that the concentration profile $C_t(x)$ for one of the elements (*e.g.,* Copper) at time $t$ of the diffusion can be expressed as a convolution product

$$C_t(x) = (C_0 \otimes G_{\sqrt{2Dt}})(x) \qquad (6)$$

where $C_0$ denotes the concentration profile of the considered element without diffusion (*e.g.,* two step functions as in Eqn. (5) in the case of the Ni/Cu/Ni material). This expression clearly shows the similarity between the measurement artefact and a diffusion equation. The raw measurement of the concentration profile is given by

$$M(x) = (C_t \otimes G_\sigma)(x) \qquad (7)$$

By combining Eqns. (6) and (7), the measured profile can be expressed as

$$M(x) = (C_0 \otimes G_\Sigma)(x) \qquad (8)$$

with

$$\Sigma = \sqrt{\sigma^2 + 2Dt} \tag{9}$$

Equation (8) illustrates the artefact created by the *pear effect* on the diffusion coefficient since the apparent one $D^*$ becomes $D^* = D + \sigma^2/2t$. We can see that this artefact is all the more important when the ratio $2Dt/\sigma^2$ is small compared to 1. An example is presented in Fig. 5 for the multilayered material where the ratio $2Dt/\sigma^2 = 0.1$. In this case, the normalised true diffusion length $k_t$, which is expressed as $k_t = \sqrt{2Dt} / h_{Cu}$, is equal to 0.1 while the apparent one $k_a$ is equal to 0.54. Figure 6 shows the change in error between the true and apparent diffusion lengths related with the true diffusion length. In our case, the true diffusion length is generally low, therefore it is necessary to deconvolute the apparent diffusion coefficient by using Eqn. (9) since $\sigma$ and $t$ are known.

Conclusions:

The study described herein focuses on a non-standard artefact that is always present in an EDX measurement of concentration profiles in an SEM. It is called *pear effect* because it finds its origin in the shape of the X-ray emission volume. The artefact on the study of diffusion profiles has been completely developed in the case of close atomic number elements for the diffusion couple and a practical deconvolution *strategy* has been deduced. The importance of this artefact for the present scale of observation and diffusion can be summarised by Fig. 6 as the error tends to infinity for very low diffusion lengths. Conversely, for high diffusion lengths, the relative error vanishes and the artefact can be ignored since, for the EDX technique, a maximum mass resolution of about 1% is obtained in favourable conditions.

Acknowledgements:

   The authors wish to thank Mr. Duval of AER Company for providing the samples.

References:

1. Heinrich, K.F.J. In: *Electron beam X-ray microanalysis*, Eds: Von Nostrand Reinhold Co., New York, USA. 1981.

2. Borzenko, A.G., *et al*. Reference materials for quantitative microprobe analysis. *Microscopy and Analysis*, European edition, **62**, 29-31, November 1999.

3. Wittry, D.B. Metallurgical applications of electron probe analysis. *Advances in X-ray analysis* **3**, 197-212, 1960.

4. EDX Microprobe PGT. Integrated microanalyser for imaging and X-ray. Princeton Gamma Tech. 1994.

5. Garden, J. Confort dans l'utilisation des détecteurs de rayons X. *Journées des utilisateurs EDS/PGT – Synergie 4*, Université d'Evry, France, 1999 (in French).

6. Philibert, J. Chapter I: Diffusion and drift. In: *Atom movements - diffusion and mass transport in solids*. Eds: Les éditions de la Physique, les Ulis, France. pp. 1-28, 1996.

7. Arnould, O., *et al*. Identification d'un mode de vieillissement par diffusion d'un assemblage Nickel/Cuivre. *Groupe de travail MECAMAT "Couplages Mécaniques et Diffusion"*, École des Mines de Nancy, France. November 1999 (in French).

8. Heinrich, K.F.J. Use of Monte Carlo calculations in electron probe microanalysis and scanning electron microscopy. *Proceedings of a workshop on Electron Microscopy held at the NBS*, Gaithersburg, USA. 1976.

9. Curgenven, L., *et al*. TIRL No. 303, 1971. See also Maurice, F. Chapitre 4 : Emission X. In: *Microanalyse et microscopie électronique à balayage*. Eds: Les éditions de la Physique, Orsay, France (in French).

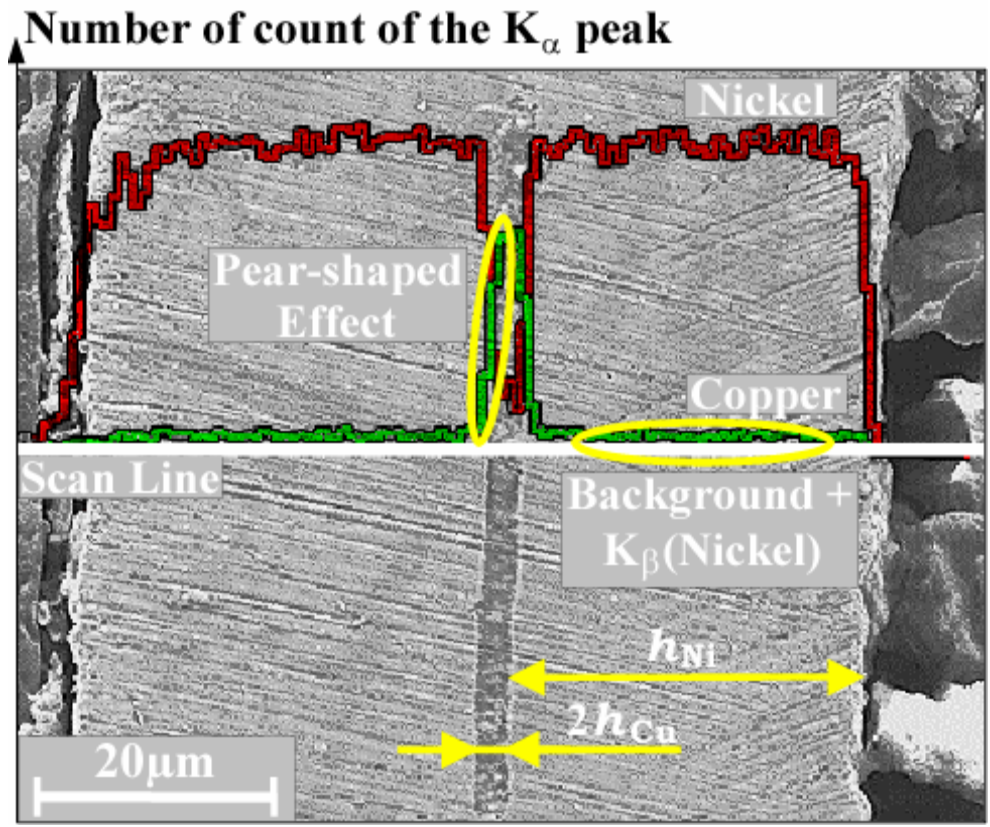

Figure 1: Concentration profile of a non-aged Ni/Cu/Ni bellows and measurement artefacts.

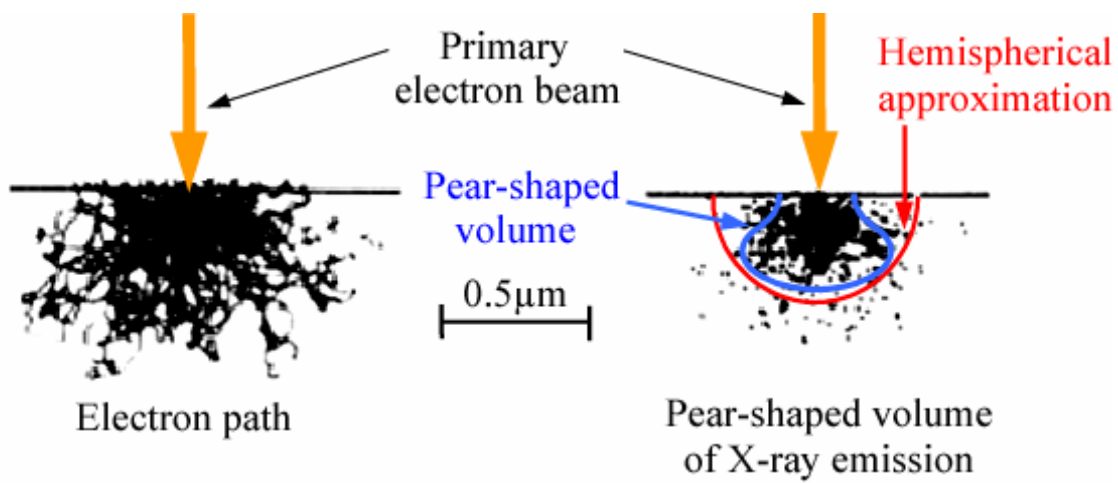

Figure 2: Electron trajectories and X-ray emission volume in Copper (acceleration voltage: 20keV) obtained by Monte-Carlo simulations [9].

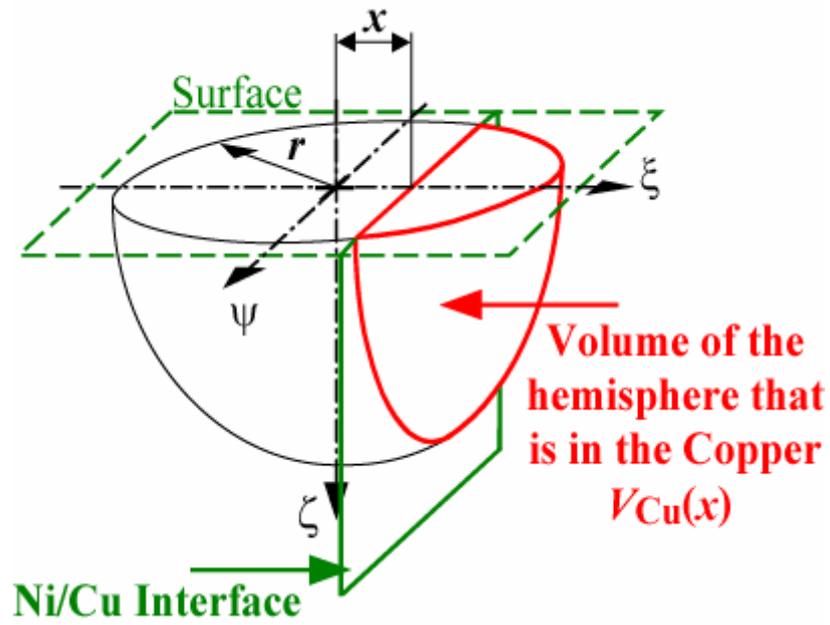

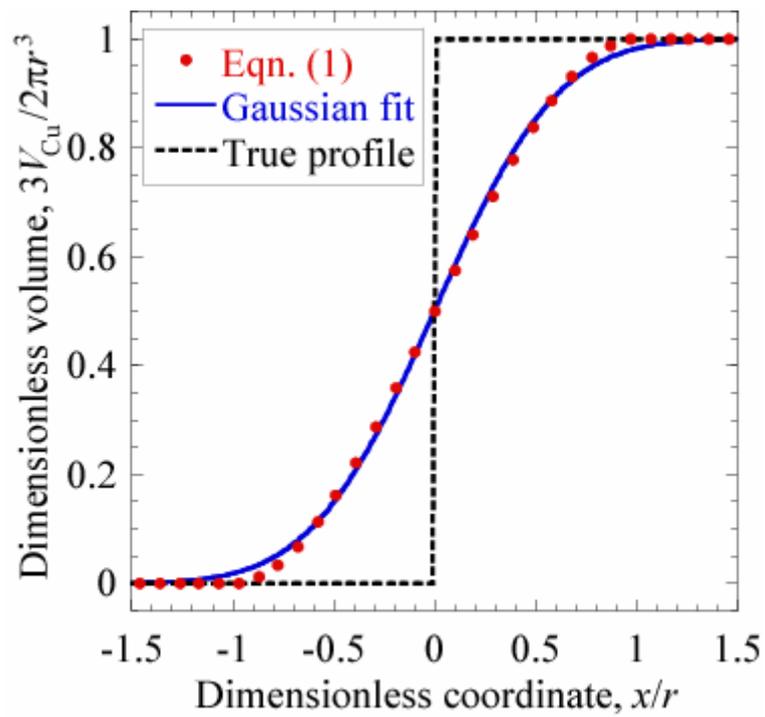

Figure 3:  (a) Depiction of $V_{Cu}$.

(b) Change in $V_{Cu}$ with the dimensionless coordinate $x/r$.

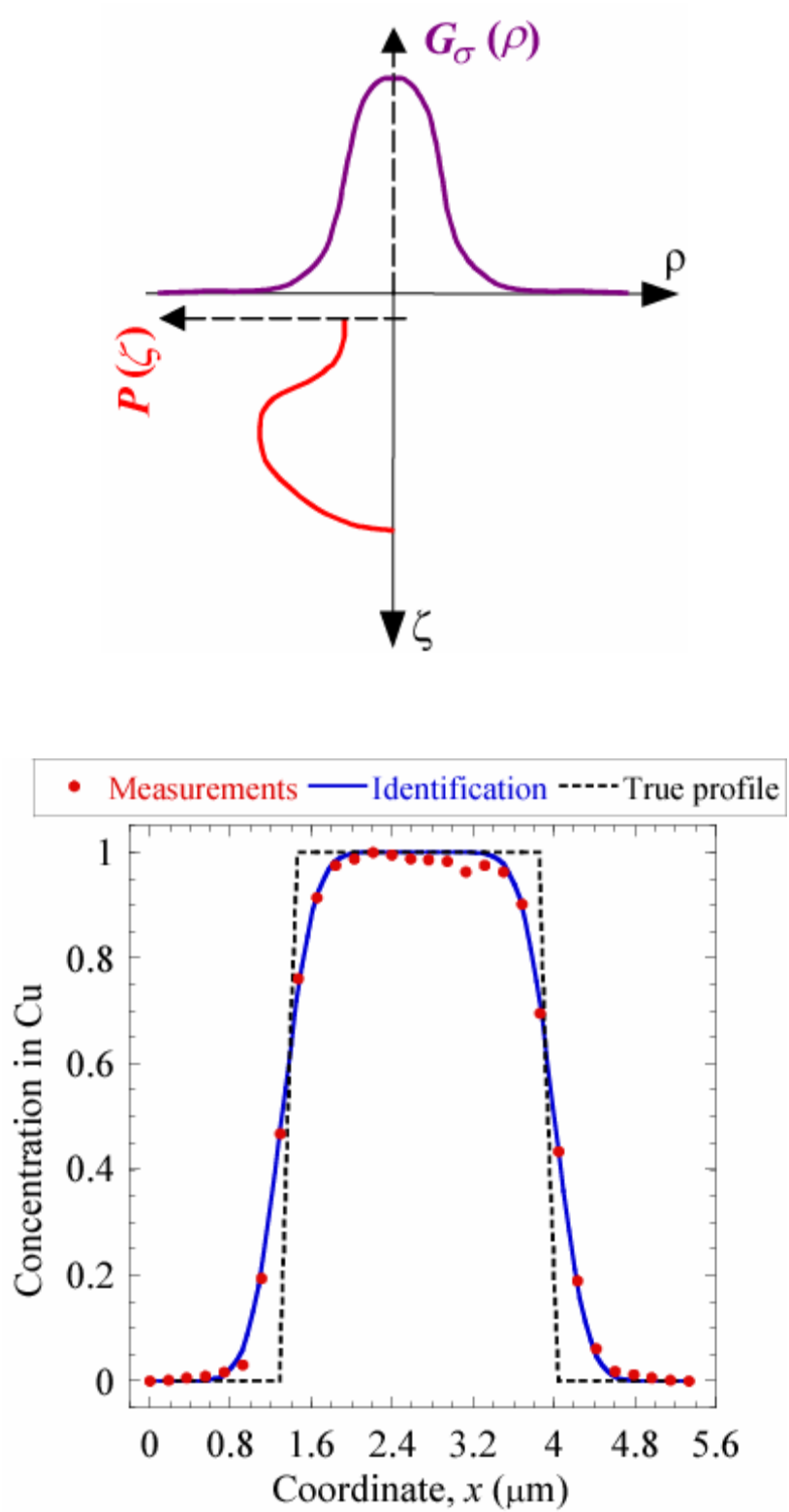

Figure 4: (a) Depiction of the emission function *E*.

(b) Identification of $\sigma$ on a reference Ni/Cu/Ni material.

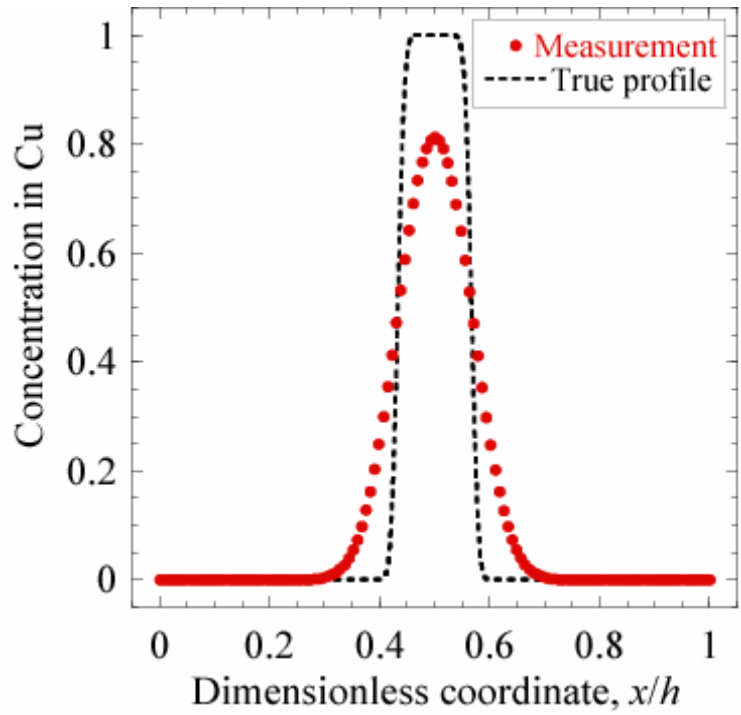

Figure 5: Comparison of a raw measurement and the actual concentration profile for $\sigma/h_{\text{Cu}} = 1.34$ and $k_{\text{t}} = 0.1$ ($h = 10h_{\text{Cu}}$).

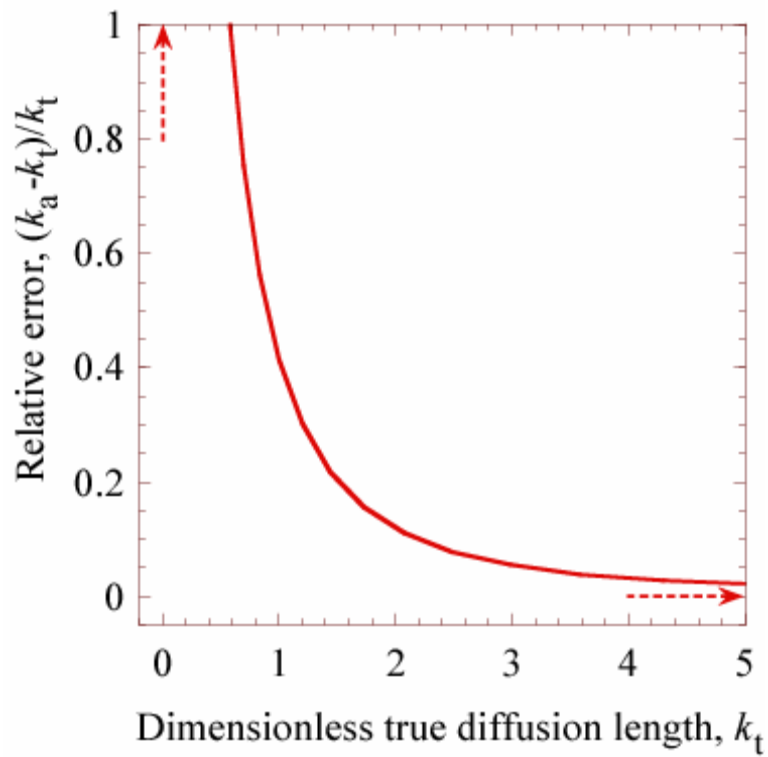

Figure 6: Relative error between the apparent ($k_{\text{a}}$) and true ($k_{\text{t}}$) diffusion length vs. the true diffusion length ($k_{\text{t}}$).